\documentclass[10pt,conference]{IEEEtran}

\usepackage{xspace}
\usepackage{algorithm}
\usepackage{algorithmic}
\usepackage{graphicx}
\usepackage{textcomp}
\usepackage{multirow}
\usepackage{tcolorbox}
\usepackage{xfp}
\usepackage{url}
\usepackage{xcolor}
\usepackage{hyperref}

\newcommand{\tool}{\textsc{\textsc{CValue}}\xspace}

\IEEEoverridecommandlockouts
\begin{document}

\title{Who is the Real Hero? Measuring Developer Contribution via Multi-dimensional Data Integration}

\author{
\IEEEauthorblockN{Yuqiang Sun\IEEEauthorrefmark{1},
Zhengzi Xu\IEEEauthorrefmark{1}\IEEEauthorrefmark{4},
Chengwei Liu\IEEEauthorrefmark{1},
Yiran Zhang\IEEEauthorrefmark{1},
Yang Liu\IEEEauthorrefmark{1}
}
\IEEEauthorblockA{suny0056@e.ntu.edu.sg, zhengzi.xu@ntu.edu.sg}
\thanks{\IEEEauthorrefmark{4} Zhengzi Xu is the corresponding author.}
\IEEEauthorblockA{\IEEEauthorrefmark{1}School of Computer Science and Engineering, Nanyang Technological University, Singapore}

}

\maketitle

\DeclareRobustCommand{\answerbox}[2][gray!20]{%
\begin{tcolorbox}[
        % breakable,
        left=0pt,
        right=0pt,
        top=0pt,
        bottom=0pt,
        colback=#1,
        colframe=#1,
        width=\linewidth, 
        enlarge left by=0mm,
        boxsep=5pt,
        arc=0pt,outer arc=0pt,
        ]
        #2
\end{tcolorbox}
}

\begin{abstract}
% \td{need to be rewritten}
%The development and maintenance of large software is often the result of teamwork. In teamwork, a method is needed to distinguish the contributions of different developers. This approach needs to be accurate, and fair and take different development scenarios into account to motivate team members. 
% It is important to measure the individual code contribution in software development activities since it helps to motivate the developers who make worthy commitments.
% % \zz{to motivate} 
% %To discover the core maintenance team of the project, as well as to motivate the developers who made real contributions, an accurate developer contribution measurement method is needed.
% Existing approaches count the number of line changes in the commits as the developer contribution. However, the feature is not robust since the bulk addition and deletion of unimportant code can easily receive a high contribution score.
% Other works leverage business value or code quality changes as indicators to measure the contribution. Unfortunately, these features cannot be always available in software projects, which makes them difficult to be applied to real-world contribution measurement.

Proper incentives are important for motivating developers in open-source communities, which is crucial for maintaining the development of open-source software healthy.
%proper incentive -> motivate developer
% Developers who contribute more should receive more rewards and recognition. 
To provide such incentives, an accurate and objective developer contribution measurement method is needed.
% to motivate -> need a measurement
However, existing methods rely heavily on manual peer review, lacking objectivity and transparency. 
The metrics of some automated works about effort estimation use only syntax-level or even text-level information, such as changed lines of code, which lack robustness.
% 描述一下，让人觉得太简单了
Furthermore, some works about identifying core developers provide only a qualitative understanding without a quantitative score or have some project-specific parameters, which makes them not practical in real-world projects.

To this end, we propose \tool, a multidimensional information fusion-based approach to measure developer contributions.
\tool extracts both syntax and semantic information from the source code changes in four dimensions: modification amount, understandability, inter-function and intra-function impact of modification. 
It fuses the information to produce the contribution score for each of the commits in the projects.
Experimental results show that \tool outperforms other approaches by 19.59\% on 10 real-world projects with manually labeled ground truth.
%By fusing the information, we achieve a $19.59\%$ improvement compared to the line-of-code-based approach. 
% Although it contains time-consuming program analysis processes,
We validated and proved that the performance of \tool, which takes 83.39 seconds per commit, is acceptable to be applied in real-world projects. 
Furthermore, we performed a large-scale experiment on 174 projects and detected 2,282 developers having inflated commits. 
Of these, 2,050 developers did not make any syntax contribution; and 103 were identified as bots.
\end{abstract}  
\begin{IEEEkeywords}
Open-Source Incentive, Mining Software Repositories, Program Analysis.
\end{IEEEkeywords}
\section{Introduction}

The Internet is dominated by open-source software, from end-user to server-side~\cite{howell2022decentralized}. 
These open-source projects, have become the foundation of widely used modern software, and are developed and maintained by passionate, yet often unpaid, developers.
Effective open-source development requires cooperation and recognition of the contribution of each developer~\cite{Zhang_Liu_Xu_Chen_Fan_Zhao_Wu_Liu_2023, zhang2023software}. 
% web3 projects start to provide decentralized ... 
% web3 organizations such as TEA & gitcoin start to -> motivate -> open-source developers -> lack of proper methods
% \yq{Financial support is a common way to encourage developers, proportional to their contribution.}
Web3 projects, such as TEA~\cite{teaxyz} and GitCoin~\cite{gitcoin}, started to realize the importance of motivating developers in open-source communities.
% Recent efforts, such as Tea~\cite{teaxyz} and GitCoin~\cite{gitcoin}, aim to allocate grants from investors to open-source projects and developers through blockchain-based methods for fairness and transparency. 
However, they lack proper methods to measure the contribution of each collaborator.
Existing measurement methods mainly rely on peer review, and lack objectivity and transparency, contravening the principles of Web3 projects. 
An unfair allocation may significantly decrease efficiency and harm developer morale.
Therefore, an objective method to measure the value of each developer who contributed to the project is necessary~\cite{shull2007guide,herbsleb2003formulation}.

No existing study focuses on measuring the value provided by developers in software, but there are some similar topics. 
Several studies about developing effort estimation methods, such as COCOMA~\cite{boehm1995cost,boehm2008achievements} and its variants, estimate the effort needed to implement specific functionality. 
However, these methods only consider code at the syntax or text level, lacking a deeper understanding of the semantics of modified code segments, and failing to pay attention to the difference between different types of modifications.
There are other studies that focus on identifying core developers within a team. 
For example, studies such as Mockus et al.~\cite{mockus2002two} and Bella et al.~\cite{di2013multivariate} focus on identifying core developers based on metrics like lines of code, commits, complexity, and comment modifications. 
Joblin et al.~\cite{joblin_classifying_2017} and Cheng et al.~\cite{cheng2019activity} propose network-based and activity-based approaches, respectively. 
These studies focus on classifying developers into different categories rather than quantifying their contribution, which could not be applied to solve the measurement problem.
A study by Tasy et al.~\cite{tsay_influence_2014} calculates the effort in pull requests from GitHub, considering the communication between developers in open-source communities.
Some of the data used in these researches do not commonly exist in open-source projects or require domain knowledge to understand and analyze, such as pull requests from GitHub, file modification other than the code, etc.
Considering the previously mentioned issues, developing a robust code-based developer contribution metric algorithm for benefit allocation faces several challenges. 
First, the analysis should base on code changes that exist in all kinds of projects.
Second, the analysis approach should focus on the changes made at the syntax level, where the quantity of modifications directly reflects the contribution of developers. 
Third, the analysis should also consider the semantic level, since syntax-level changes alone are not enough to accurately reflect the value a developer adds to the project. 
For instance, formatting and moving code provide few semantic changes, which contribute little to the project, but adding features or fixing bugs may cause semantic breaking~\cite{Zhang_Liu_Xu_Chen_Fan_Chen_Liu_2022, zhang2023mitigating}.
Fourth, the analysis method must take into account the interaction of the changed part and the whole program, which is important for developers to understand the project and make proper modifications.
To address these challenges, we propose \textbf{C}ontribution\textbf{Value}, a method that integrates code-based multi-dimensional features.
\tool leverages four types of information to assess code changes from different perspectives and produce a comprehensive score to measure the extent to which the developer contributes valuable code to the project.
For the first and second challenges, \tool uses AST differences to measure syntax-level changes in code, and exclude some non-semantic change patterns, such as changing the name of variables.
For the third challenge, \tool takes into account complexity metrics, including cyclomatic complexity, lines of code, Halstead volume~\cite{halstead1977elements}, and percentage of comments, to measure the effort required to understand the existing code.
% Why integrity?
For the fourth challenge, a call graph is built, and the position of the changed code segment in the call graph is used to represent its interaction with the whole project, and furthering measure the inter-function consideration of the developer when modifying the code.
\tool also traces the data flow and control flow of the variables involved in the change, evaluating the range of context that should be considered by developers when making modifications to the code segment.
In the end, \tool integrates information from these four aspects and provides a more comprehensive view of the change from the code level.
% \td{MAP CHALLENGE TO SOLUTION}
% TODO: Contribution
% Should be changed with the evaluation part.
% To evaluate \tool, we manually labeled 1398 commits on 10 popular projects under Google OSS-Fuzz project~\cite{oss-fuzz} to construct the ground truth. The experimental result shows that \tool outperforms the baseline methods with the spearman correlation of $19.59\%$. Moreover, future analysis has shown that \tool produces a more reasonable assessment than human-labeled ground truth on 1398 commits. On the given project, the average time spent on each commit during \tool analysis was 68.24 seconds on average with a maximum of 121.04 seconds.

% In this paper, we made the following contribution:
% \begin{enumerate}
%     \item We produce \tool, a framework to quantify the contribution made by developers in code commits.
%     \item We build an incremental dependency extraction tool, which optimized the time cost for analysis on git repositories.
%     \item We conduct an empirical study with \tool on 174 open source projects and found 2282 developers with significant gaps in their number of commits and \tool to the project.
% \end{enumerate}

To evaluate the effectiveness of \tool, we manually labeled 1,398 commits on 10 popular projects under the Google OSS-Fuzz project~\cite{oss-fuzz} to construct the ground truth. The experimental results indicated that \tool outperformed the baseline methods, with a Spearman correlation of 19.59\%. 
% Furthermore, the future analysis revealed that \tool produces a more reasonable assessment than human-labeled ground truth on the 1398 commits. 
On the given project, the average time spent on each commit during the \tool analysis was 83.39 seconds, with a maximum of 121.04 seconds.

In summary, the main contributions of this paper are:
\begin{enumerate}
    \item We proposed \tool, a framework for quantifying the contribution made by developers in code commits.
    \item We built an incremental dependency extraction tool, which optimized the time cost for analysis on git repositories.
    \item We conducted an empirical study with \tool on 174 open source projects and found 2,282 developers with significant gaps in their number of commits and \tool to the project.
    \item Our data and artifact are published on the website\footnote{\url{https://sites.google.com/view/ase23contributionmeasurement}}.
\end{enumerate}
 
\section{Related Works}
% \td{LOOKS LIKE RELATED WORKS}

% \subsection{Related Works}
% There is no existing work to measure the value developers contribute to a project. There are two main types of research that are similar. 
% One is the estimation of developing effort, which is time spent on completing a particular development task; the other is the finding for core, in other words, more valuable, developers.
% 第一段直接一点
Previous research primarily focused on estimating the effort required for the completion of development tasks or identifying the key or valuable developers within a project, which cannot be applied to contribution measurement and motivating developers. 
There was a lack of studies specifically measuring the overall value that developers contribute to a project.

One similar topic, effort estimation, has been widely researched. 
One commonly used method for effort estimation is the COCOMO model~\cite{boehm1995cost,boehm2008achievements}, which is based on a regression analysis of lines of code. 
Several studies have attempted to improve the accuracy of the COCOMO model by incorporating additional information, such as neural networks~\cite{huang2007improving,sachan2016optimizing,nguyen2008constrained}. However, they are not suitable for large-scale analysis as the parameters of the model need to be determined by the project manager of each project. 
Furthermore, it has been established that the number of lines of code modified alone is not a reliable indicator of developer contributions~\cite{shihab2013lines,lima2015assessing}.

Recently, Yin et al.~\cite{Yin:EECS-2018-174} and Ren et al.~\cite{ren2018towards} proposed an approach to measure developer contributions using a call-graph-based DevRank algorithm. 
This approach uses not only the number of lines of code modified but also the position of the code in the call graph and the type of modification. 
Natural Language Processing (NLP) techniques were applied to process commit messages in order to distinguish different types of commits and assign different weights to different types of modifications. 
The final contribution value was obtained by combining these two metrics using a Learning to Rank (L2R)~\cite{liu2009learning} method. 
However, the authors did not address the issue of determining the weights for different types of modifications and whether the parameters of the L2R method are shared across projects.

Previous research on identifying core developers in software development projects has employed a variety of methods. 
Mockus et al.~\cite{mockus2002two} used the number of lines of code committed by each developer as the sole criterion for classification. 
Bella et al.~\cite{di2013multivariate} employed a multivariate analysis that considered lines of code, number of commits, complexity, comment modifications, structural changes, and non-structural changes. 
Jergensen et al.~\cite{jergensen2011onion} only considered the activities of developers, such as the number of lines of code added, which may not accurately reflect their role in the project.
Joblin et al.~\cite{joblin_classifying_2017} proposed a network-based approach that took into account cooperation in version control systems and email networks. 
Cheng et al.~\cite{cheng2019activity} and Izquierdo et al.~\cite{canovas2022analysis} used activity-based role detection methods, by considering code contribution, opinion contribution, network interactions, and administration roles. 
{\c{C}}etin et al.~\cite{ccetin2022analyzing} used the change history of each file to determine the corresponding developer and divided the developers in the community into three roles, including Jacks, Mavens, and Connectors. 
Robles et al.~\cite{robles2009evolution} used the frequency of committing code in a certain period, Bella et al.~\cite{di2013multivariate} used the number of lines of code, the number of files, complexity (McCabe cyclomatic complexity~\cite{mccabe1976complexity}), and structure/non-structure modifications, etc. to classify developers into different roles.
However, these methods treat identifying core developers as a classification problem rather than quantifying the contribution of each individual and are therefore not suitable for this allocation problem.

\section{Methodology}

\begin{figure*}
    \centering
    \includegraphics[width = \linewidth]{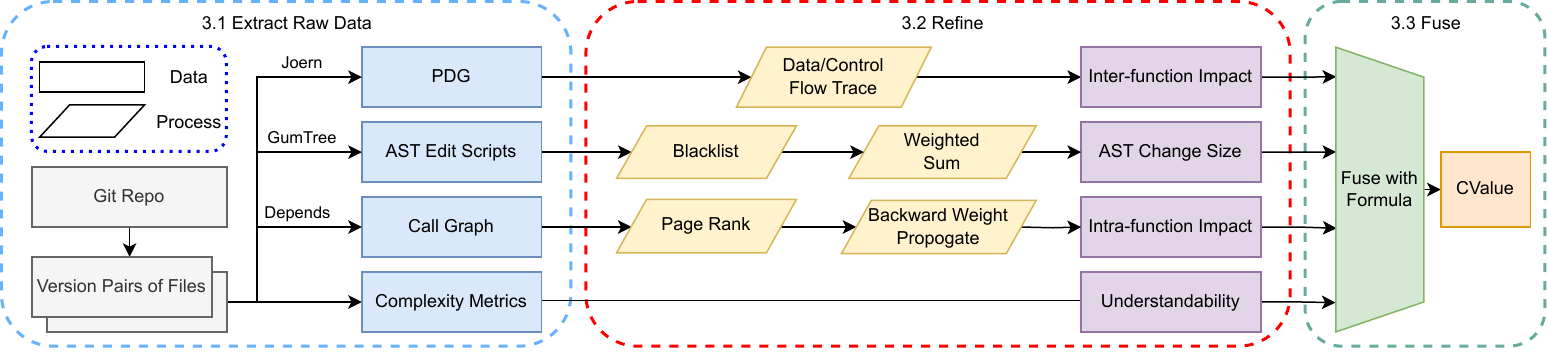}
    \caption{Overview of \tool}
    \label{fig:overview}
\end{figure*}
% 加一点目的性的内容，为什么要这么做

Figure~\ref{fig:overview} illustrates the overview of \tool, which have three steps.
% total 3 steps
% \td{
% As shown in the figure, \tool extracts a series of features from code changes, which are then processed to represent four dimensions of information.} 
% Then, we use a formula to aggregate all the data and obtain a score that measures the amount of contribution.
In the first step, four kinds of information are extracted from the files, which are program dependence graph (PDG), abstract syntax tree (AST) edit scripts, call graph, and complexity metrics.
In the second step, we process the extracted raw data, converting them into a series of values that reflect the considerations of developers from four perspectives when modifying the code, which are the AST change size, understandability, inter-function, and intra-function impact of the changed code segment.
Finally, we fuse the values from the four perspectives into a single value, used to measure the contribution of developers.

% The proposed methodology for this research is outlined in three steps. 
% Firstly, various types of information are extracted from source code in Git repositories, including AST difference, complexity metrics, call graph, and program dependence graph (PDG). 
% In the second step, this extracted information is refined to focus on specific aspects, such as the size of the change, understandability, the inter-function and intra-functino impact when modifying the code. 
% Finally, all of the refined data is fused together to create a score that measures the value a developer contributes to the project. 
% A diagram illustrating the overview of \tool can be seen in Figure~\ref{fig:overview}.
% The overview of \tool is shown in Figure~\ref{fig:overview}. 
% The method can be summarized in three steps.
% First, four types of information are extracted from source code in Git repositories, including AST difference, complexity metrics, call graph, and program dependence graph (PDG). 
% In the second step, we refined the extracted information to be the size of the change, readability, the interaction of the whole program, and intra-function concer ns when modifying the code.
% Finally, we fused all these refined data to a score to measure the value a developer contributes to the project. 

\subsection{Data Extraction}

% We extracted four dimensions of raw information from the code to measure the value contributed by the programmers.
% The dimensions consist of AST difference, complexity metric, call graph, and program dependence graph.
% We will discuss the extraction procedure of the four dimensions in the following section.
\tool involves extracting four dimensions of raw information from the code to accurately measure the value contributed by programmers. 
These dimensions include AST difference, complexity metrics, call graph, and program dependence graph. 
The procedures for extracting these dimensions will be discussed in detail in the following section of the paper.

\subsubsection{AST Difference - Syntax Level Change} 

%In the proposed method, we 
% \tool uses the difference of AST by comparing the tree of files from two versions to quantify the amount of changed code. 
% AST provides a view of code from a syntax level. 
% Using this metric, changes that do not produce syntax changes, such as adjusting formatting, modifying comments, and changing variable names, can be effectively filtered. 
% It also distinguishes different types of modifications such as additions and deletions.
One of the dimensions used in \tool is the AST difference, which quantifies the amount of code that has been changed. 
This is achieved by comparing the tree of files from two versions. 
The use of AST (Abstract Syntax Tree) provides a syntax-level view of the code. 
This metric allows us to effectively filter out changes that do not produce syntax changes, such as adjustments of code format, modifications to comments, and changes of variable names. 
Additionally, it allows us to distinguish different types of modifications, such as additions and deletions of AST nodes.

% In this part, we used the mapping policy of GumTree~\cite{gumtree}. 
% GumTree generates the difference between two source code files and edit scripts which describes how to change from one to another. 
% After extraction, we got the edit script, which lets us know the exact content of each change. 
% Finally, we counted different types of modification operations grouped by each method for subsequent analysis.
To extract the AST difference, we use the mapping policy of GumTree~\cite{gumtree}. 
GumTree generates the difference and edit scripts between two source code files, which describe how to change from one version to another. 
After the extraction, we obtain the edit script, which provides us with the exact content of each change. 
We count the different types of modification operations grouped by each method for subsequent analysis.

\subsubsection{Complexity - Readability}

% We used complexity metrics to reflect the thinking when a developer read the specified code segmentation. 
% A more complex code is rather difficult to be understood and maintained. 
% Therefore, developers should put more effort into making changes to complex code.
% The complexity metrics are measured at the method level. 
% These metrics include four parts: 
% 1) Line of Code of a function. 
% A function with a larger number of code lines is harder to comprehend. 
% 2) Halstead Volume~\cite{1977Halstead}. 
% Line of Code does not consider the length of each line. 
% Halstead Volume is based on the number of operators and operands, which prevents the calculation method from being attacked simply by increasing the number of lines of code. 
% 3) Percentage of comments. 
% Comments help developers understand the logic of the source code. 
% 4) Cyclomatic Complexity~\cite{mccabe1976complexity}. 
% Cyclomatic Complexity (CC) is a measure of the possible execution paths of a function.
% A larger CC means that the function has more possible execution paths, and function is more complex and difficult to understand.

To reflect the effort required to read and understand a specific code segment, we use method-level complexity metrics. 
A more complex code is harder to understand and maintain, and thus, developers should put more effort into making changes to such code, and \tool will give a higher weight to code with higher complexity. 
The complexity metrics include four parts:
1) Lines of Code of a function. 
A function with a larger number of code lines is often harder to comprehend.
2) Halstead Volume~\cite{1977Halstead}. 
Line of Code does not consider the length of each line. 
Halstead Volume is based on the number of operators and operands, which prevents the calculation method from being attacked simply by increasing the number of lines of code.
3) Percentage of comments. 
Comments help developers understand the logic of the source code.
4) Cyclomatic Complexity~\cite{mccabe1976complexity}. 
Cyclomatic Complexity (CC) is a measure of the possible execution paths of a function. 
A larger CC means that the function has more possible execution paths, and thus, the function is more complex and difficult to understand.

% 不要用Integrity
\subsubsection{Call Graph - Inter-function impact of changes}

% We should the integrity of the project, i.e. the changed code segment may interact with the rest part, and these relations should be considered.
% % We want to determine the importance of different methods by calculating the dependencies between methods. 
% The call graph presents a kind of dependency in the given program. 
% When a developer modifies code in a project, it should consider how the code segments are being called or how the code segments invoke other methods in the project. 
% The call graph was extracted by Depends~\cite{multilangdependsdepends}, a dependencies extraction tool for multi languages based on source code. 
% After obtaining the call graph, we will describe how to calculate the call graph into the importance of methods in the next section.

\begin{figure}
    \centering
    \includegraphics[width=\linewidth]{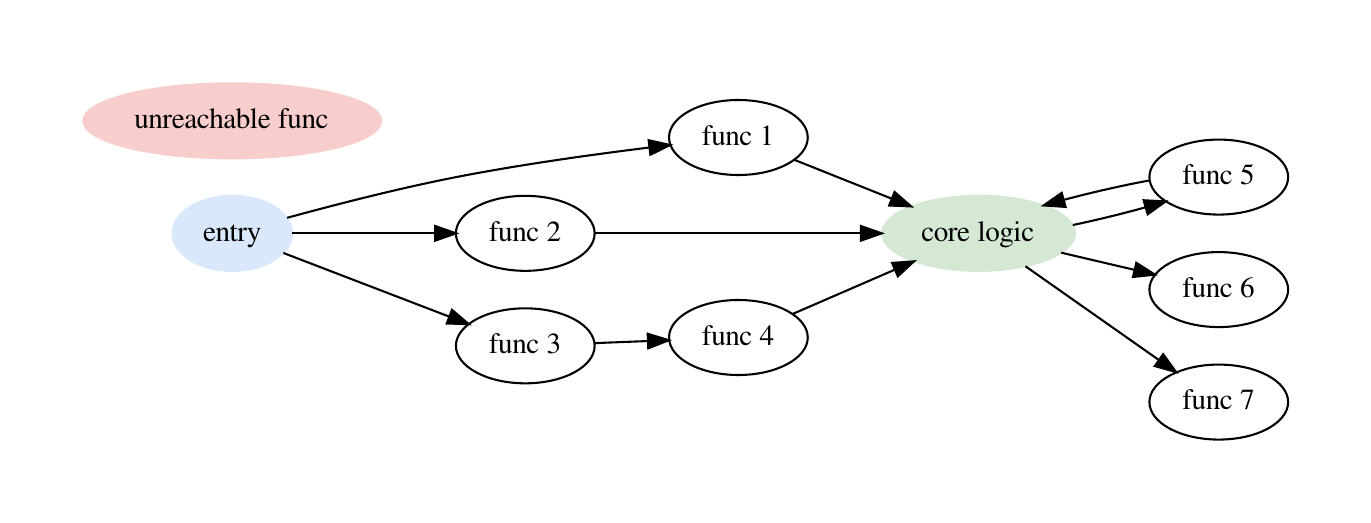}
    \caption{Call graph example}
    \label{fig:callgraph}
\end{figure}

% To consider the cohesion of the project, we also analyzed the interactions between the changed code segment and the rest of the program. 
Considering the cohesion among various components is crucial for programmers when making modifications, as software is an integrated entity composed of different parts.
Developers need to exercise greater caution when editing code at core positions, thus \tool assigns a higher weight to such modifications.
For example, as shown in Figure~\ref{fig:callgraph}, \tool assigns a higher weight to developers who make modifications to the core logic.
Conversely, developers who make modifications to less important code segments or even unreachable code will be assigned a lower weight by \tool.
% 改配色
To do this, we use the call graph, which presents a kind of dependency in the given program. 
% When a developer modifies code in a project, it is important to consider how the code segments are being called or how the code segments invoke other methods in the project. 
The call graph is extracted using Depends~\cite{multilangdependsdepends}, a tool for extracting dependencies in multiple languages based on source code. 
We will further describe how to calculate the call graph into the inter-function interaction of the modified code segment in the next section.

% However, since extracting the call graph is time-consuming, we used an incremental dependency extraction tool based on Depends to get the call graph of each version. 
% In this incremental analysis approach, we build the call graph sequentially based on the software version tree in the Git repository using a depth-first traversal method. 
% When the software version changes, only the dependencies involved in the modified file are updated, other unmodified file-related dependencies are not updated. 
% When there is a fork in the software version tree, the dependency extraction tool caches the version before the fork. 
% When the analysis of one branch is finished, it restores the cached results and analyzes another branch.
However, since extracting the call graph could be a time-consuming task, we used an incremental dependency extraction tool based on Depends to obtain the call graph for each version. 
This incremental analysis approach builds the call graph sequentially based on the software version tree in the Git repository using a depth-first traversal method. 
When the software version changes, only the dependencies involved in the modified file are updated, while other dependencies related to unmodified files are not updated. 
Additionally, when there is a fork in the software version tree, the dependency extraction tool caches the version before the fork. 
Once the analysis of one branch is finished, it restores the cached results and begins the analysis of another branch.

\subsubsection{PDG - Intra-function impact of changes}

\begin{figure}
    \centering
    \includegraphics[width=\linewidth]{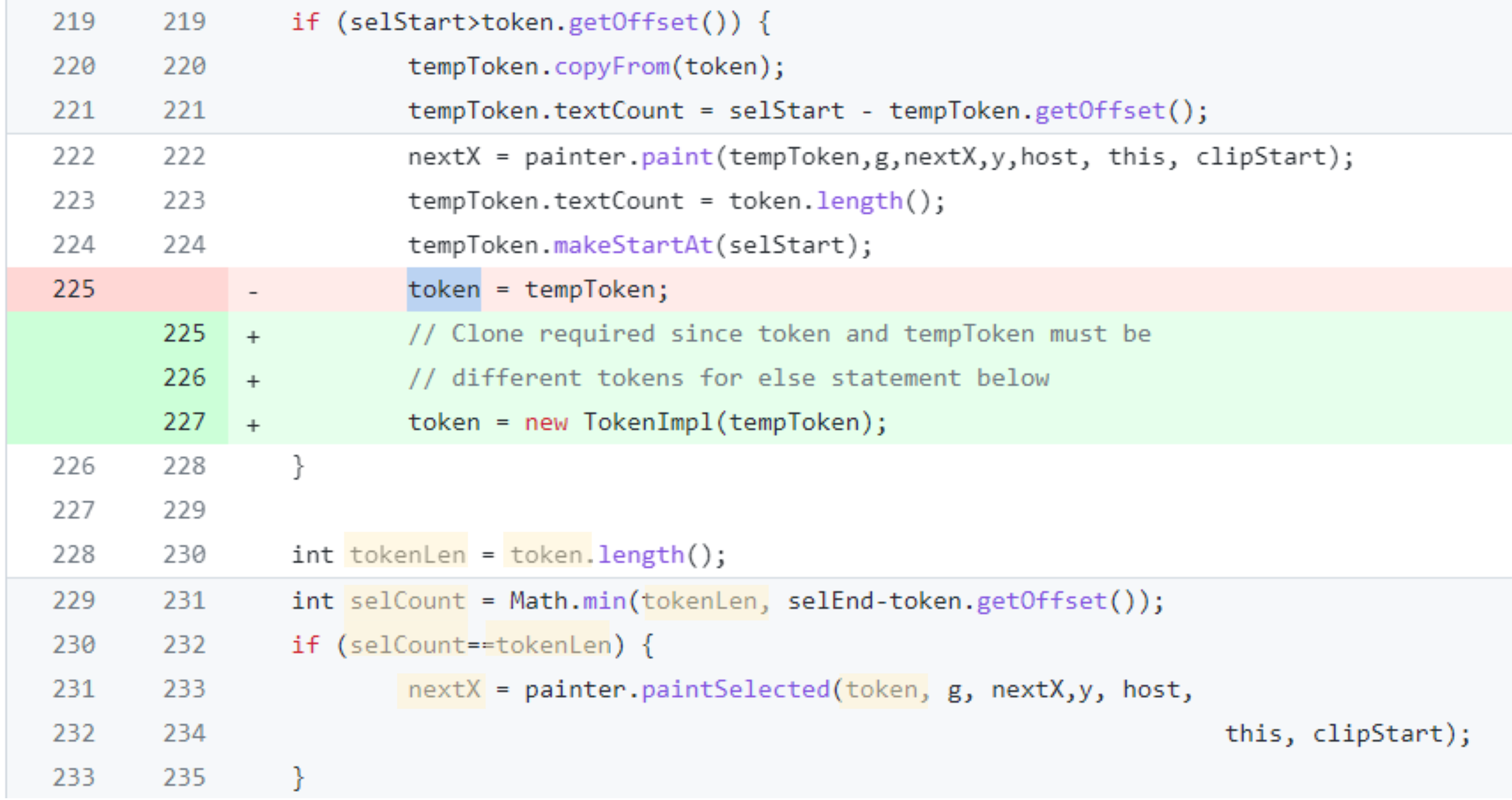}
    \caption{Intra-function impact example}
    \label{fig:pdg}
\end{figure}

% The inter-functional call relationship is not the only thing that needs to be considered when making changes to code.
% Apart from the importance of the method, the contribution value of different changes in the same method may vary.
% % \zz{NOT CLEAR}
% Different variables, or conditional statements, have different scopes of influence within the same method. More care needs to be taken when making changes to those variables and statements that have a more significant impact on existing code.
% % \zz{HOW IT AFFECT THE CONTRIBUTION?}
% \tool used PDG, which integrates the data flow dependencies and control flow dependencies of program fragments to capture the intra-function influences of the changes. 
% % \zz{ADD EXAMPLES}
% For example, the contribution of modifying an expression in a complex conditional statement should be different from modifying a variable initialization statement.
% \tool extracted PDG from the project source code using  Joern~\cite{yamaguchi2014modeling, joern}, a source code static analysis tool.
In addition to analyzing the connection among methods in the project, it is also important to consider how the contribution value of different changes varies even within the same method. 
Different variables or conditional statements have different scopes of influence within the same method, and thus, \tool will give higher weights for changes to those variables and statements that have a more significant impact on existing code.
As shown in Figure~\ref{fig:pdg}, the change modified line 225 to 227, and variable \textit{token} is influenced, which may have an impact on a series of variables after the changed lines (marked yellow in the figure).
To capture these intra-function influences of the changes, our tool uses the program dependence graph (PDG), which contains the data flow dependencies and control flow dependencies of program fragments. 
For example, the contribution of modifying an expression in a complex conditional statement should be different from modifying a variable initialization statement. 
We extracted the PDG from the project source code using Joern~\cite{yamaguchi2014modeling, joern}, a source code static analysis tool.

% \td{CAN ADD IN CODE EXAMPLES HERE}

% Note that \tool was designed to measure the programmer contribution in the submitted code. 
% Therefore, all types of information in this method are derived from the code and its changes. 
% Contributions beyond the code, such as communication with other developers about the requirements and design of the program in Pull Requests, are not considered in the proposed method since such contributions cannot be measured appropriately and can be reflected by code change to some extent.
It is important to note that our tool, \tool, is designed specifically to measure the contribution of the programmer reflected in the submitted code. 
Therefore, all types of information used in this method are derived from the code and its changes. 
Contributions beyond the code, such as communication with other developers, and discussing the requirements and design of the program in Pull Requests, are not considered in the proposed method since such contributions cannot be measured appropriately and are often reflected by code changes to some extent.

\subsection{Information Refinement\label{sec:refine}}

% After extracting the raw data from the Git repositories, they need to be refined before fusion. 
% In the 4 types of information extracted, the complexity metrics do not need to be refined before normalization.
After extracting the raw data from the Git repositories, it needs to be refined before being fused together. 
Among the four types of information extracted, the complexity metrics do not need to be refined before normalization.

\subsubsection{AST Difference}

% The difference of AST is introduced to quantify the amount of changed code. 
% Compared to using the number of lines of code being modified, using AST difference is able to filter out operations that do not affect the syntax of the code, such as modifying comments and adjusting formatting. 
% However, if we only consider the number of modified AST nodes, the following problems arise: 
% 1) Different modification types produce completely different contributions. 
% The same code is added to the project, removed from the project, or replaced in a different location, producing entirely different contributions. 
% 2) There are different types of nodes in the AST, and not all have the same meaning. For example, simply changing the variable name does not produce functional changes in the program but a significant change in AST. 
% 3) The number of AST nodes grows quickly with the number of operations involved in the statement, which can affect subsequent calculations.
The difference in AST is used to quantify the amount of changed code. 
Compared to using the number of lines of code being modified, using AST difference allows us to filter out operations that do not affect the syntax of the code, such as modifying comments and adjusting format. 
However, if we only consider the number of modified AST nodes, the following problems arise:
1) Different modification types produce completely different contributions. 
The same code added to the project, removed from the project, or replaced in a different location, produces entirely different contributions.
2) There are different types of nodes in the AST, and not all have the same meaning. 
For example, simply changing the variable name does not produce functional changes in the program but has a significant impact on AST.
3) The number of AST nodes grows quickly with the number of operations involved in the statement, which would affect subsequent calculations.
To overcome these problems, we need to refine the amount of AST difference by considering the type of modification and the type of the AST node.

% To differentiate the contribution for different AST changes, \tool assigns different weights to different types of modifications. 
% First, we assumed that no difference between adding and updating AST nodes. 
% They both modify the function for the function, so we gave them the same weight.
% Second, we believe that deleting existing code requires little thought to make the decision. 
% Even when old features are replaced with new ones, the contribution value is already counted when the new features are added. 
% Therefore, we gave the deletion operation on a low weight.
% There is more to consider when moving code than when deleting code. 
% Not all code can be moved from one place to another and keep the program running stably. 
% However, moving code does not improve the program's functionality. 
% Therefore, we gave the move operation a weight between add and delete.
% The weight is shown in Table~~\ref{tab:weight_of_change_type}.
To differentiate the contribution for different AST changes, \tool assigns different weights to different types of modifications.
First, we assumed that there is no difference between adding and updating AST nodes. 
They both modify the function for the function, so we gave them the same weight.
Second, we believe that deleting existing code requires less thought to make the decision. 
Even when old features are replaced with new ones, the contribution value is already counted when the new features are added. 
Therefore, we assigned a low weight to the deletion operation, which is 1\% of adding nodes.
There is more to consider when moving code than when deleting code. 
Not all code could be moved from one place to another and keep the program running stably. 
However, moving code does not change the semantics of programs. 
Therefore, we assign a weight between add and delete to the move operation, which is 10\% of adding nodes.
% The weight is shown in Table~~\ref{tab:weight_of_change_type}.

% \begin{table}[t]
%     \centering
%     \caption{Weight of Different Types of AST Change}
%     \begin{tabular}{|c|c|c|c|}
%         \hline
%         Add & Delete & Move & Update \\ \hline
%         1 & 0.01 & 0.1 & 1 \\ \hline
%     \end{tabular}
%     \label{tab:weight_of_change_type}
% \end{table}

% We gave weights to different AST nodes to solve the second problem. 
% The first type of node is the node that represents the name of the variable. 
% The second class of nodes is modifiers, including access modifiers (such as \textit{public} and \textit{final}) and annotations (such as \textit{@Override}).
% Changing the variable name or modifiers alone does not change the functionality of the project. 
% Therefore these operations are considered a small contribution. 
To solve the second problem, we assign weights to different AST nodes.
The first type of node is the node that holds the name of the variables. 
The second type of node is modifiers, including access control modifiers (such as \textit{public} and \textit{private}) and annotations (such as \textit{@Override}).
Changing the variable name or modifiers alone does not influence the functionality of the project. 
Therefore, these operations are considered a small amount of contribution.

% To address the problem that the number of AST nodes grows too fast with statement complexity, \tool uses the depth of the modified AST subtree instead of the number of modified AST nodes. 
% Such a counting rule also discourages developers from writing overly complex code.
To address the issue of rapidly increasing numbers of Abstract Syntax Tree (AST) nodes with increasing statement complexity, \tool utilizes the depth of the modified AST subtree as a metric instead of the number of modified AST nodes. 
This counting method not only helps to mitigate the problem but also serves as a deterrent for developers to avoid writing overly complex code.
We calculate the difference by the following formula:
\begin{equation}
    \Delta_{AST} = \sum\limits_{d} type(d)\cdot depth(d)\cdot name\_change(d)
\end{equation}
where $d$ is a changed AST subtree, $type(d)$ is the weight of a specific change type, $depth(d)$ is the depth of the changed AST subtree, $name\_change(d)$ is $1$ when the changed node is not only modifier or variable name, otherwise is $0.01$.

\subsubsection{Inter-Function Impact of Changes}
% To determine the consideration of the integrity of the project from the call graph, \tool is inspired by DevRank~\cite{ren2018towards,Yin:EECS-2018-174} algorithm.
% %The position of the modified function in the call graph needs to be quantified before it can be used for subsequent computations. 
% %Jinglei Ren et al. use DevRank~\cite{ren2018towards,Yin:EECS-2018-174} to quantify the importance of a function. 
% DevRank is an improved algorithm of PageRank\cite{page1999pagerank} on the function call graph, which uses the volume of the function to replace the transition probability in PageRank.
% However, this algorithm has certain drawbacks. 
% In PageRank, the page that more pages point to will get a higher score.
% However, it is different in the function call graph. 
% The functions at the end of the call chain are often utilities, simple calculations, or judgments. 
% The critical core logic does not appear at the end or the beginning of the function call chain. 
% When modifying these utilities, developers do not need to consider too much. 
% However, when modifying the core logic, it is often necessary to consider the functionality's stability, efficiency, and usability. 
To determine the consideration of the inter-function interaction of the modified code segment from the call graph, \tool is inspired by DevRank~\cite{ren2018towards,Yin:EECS-2018-174} algorithm, which is an improved version of the PageRank algorithm~\cite{page1999pagerank} applied to the function call graph.
DevRank uses the volume of the function as a replacement for the transition probability used in PageRank. 
However, this algorithm has certain limitations. 
In PageRank, pages that are pointed to by more pages receive a higher score, but this is not always the case in the function call graph. 
Functions at the end of the call chain are often utilities, simple calculations, or judgments, while the critical core logic does not always appear at the end or the beginning of the call chain. 
When modifying these utilities, developers do not need to consider as much, but when modifying the core logic, it is important to take into account the stability, efficiency, and usability of functionalities.

% Our proposed calculation method is an improvement of DevRank.
% However, adding backward weight propagation with decay in the process allows functions located in the middle of the call chain to get higher scores. 
% Our method is shown in Algorithm ~\ref{alg:function_importance}. The call graph is a directed graph with cycles. 
% When a cycle is encountered, the weights passed in are shared with all nodes on the cycle. 
% By applying such an algorithm, the part with the highest score will not be the function at the end of the call chain but will be in the middle of the call chain.
Our proposed calculation method improves upon the DevRank algorithm by incorporating backward weight propagation with decay in the process. 
This allows for functions located in the middle of the call chain to receive higher scores. 
Our method is detailed in Algorithm ~\ref{alg:function_importance}. 
The call graph is represented as a directed graph. 
When a cycle is encountered, the weights passed in are distributed among all nodes on the cycle. 
By implementing this algorithm, the highest scoring part will not be the function at the end of the call chain, but instead will be located in the middle of the call chain.

\begin{algorithm}\small
    \caption{Measure of inter-function interaction of the modified code segment}
    \label{alg:function_importance}
    \begin{algorithmic}[1]
        \renewcommand{\algorithmicrequire}{\textbf{Input: }}
        \renewcommand{\algorithmicensure}{\textbf{Output: }}
        \REQUIRE call graph $G_{c}$, a map from node in graph to its pagerank score $map_{pr}$, and a decay value $decay\in[0,1]$.
        \ENSURE a map from nodes in the graph to its PageRank score $map_{out}$.
        
        \STATE{$map_{out} = \{\}$}
        \STATE{$map_{tmp} = \{\}$} 
        \FOR{$node$ in $G_{c}$}
            \IF{$node$ not in $map_{tmp}$}
                \STATE{process($node$, $map_{tmp}$)}
            \ENDIF
        \ENDFOR
        \FOR{node in $map_{p}$}
            \STATE{$map_{out}[node]=map_{pr}[node]+map_{tmp}[node]$}
        \ENDFOR
        \RETURN{$map_{out}$}
        \renewcommand{\algorithmicensure}{\textbf{Sub procedure 1: }}
        \ENSURE $process(node, map_{tmp})$
        \STATE{$map_{tmp}[node]=0$}
        \IF{$node.children.isEmpty()$}
            \STATE{$map_{tmp}[node]=map_{pr}[node]$}
        \ENDIF
        \FOR{$child$ in $node.children$}
            \IF{$child$ not in $map_{tmp}$}
                \STATE{$process(child, map_{tmp})$}
            \ENDIF
            \STATE{$map_{tmp}[node]+=map_{tmp}[child]\cdot decay$}
        \ENDFOR
    \end{algorithmic}
\end{algorithm}

\subsubsection{Intra-Function Impact of Changes}

% Any modification within a function involving variables may cause statements that use the variables in the context to be affected. 
% The impact range should also be considered by the developer when making a change to the code as the intra-function concern. 
% By comparing PDGs before and after revision, we know which nodes in PDG are changed. 
% The range of such effects is calculated by their data and control flow dependencies. 
When making modifications to a function that involves variables, it can potentially affect statements that utilize those variables in the context, making it important for developers to consider the impact range of their changes, as part of the intra-function concern. 
To determine the range of impact, our proposed method compares the program dependence graphs (PDGs) before and after revision to identify which nodes in the PDG have been changed. 
The range of the effects is calculated based on the data and control flow dependencies of those changed nodes.

% When calculating the affected range in the data dependency graph (DDG), we did both forward and backward traces from the changed node in the data dependence graph. 
% Finally, we counted the number of nodes involved in forward and backward tracing. 
% The ratio of the number of tracked nodes to the number of total nodes is the impact range of DDG. 
To calculate the affected range in the data dependency graph (DDG), we perform both forward and backward traces from the changed node in the DDG. 
The impact range of DDG is determined by counting the number of nodes involved in the forward and backward tracing and taking the ratio of the number of tracked nodes to the total number of nodes.

% Furthermore, we should give higher weight to logical modifications, which are modifications that involve conditional statements. 
% Even if the two modifications are similar in size, it takes more effort to modify the conditional judgment than to modify other statements. 
% Since even though there are usually only a few expressions in conditional statements, changes happening here can significantly impact the program's execution.
% Such modifications commonly happen when fixing bugs in source code. We determine whether a conditional statement has been modified by counting the number of successor nodes in the control dependence graph (CDG) of the changed node. 
% The changed node is inside a conditional statement if there is more than one successor. 
% We calculate the ratio of the number of affected nodes to the total number of nodes as the impact range of CDG.
Additionally, we assign a higher weight to logical modifications, which involve changes to conditional statements. 
Some types of modifications take more effort to make, even if they are similar in size, as changes to conditional statements can significantly impact the execution of programs. 
These modifications are commonly made when fixing bugs in source code. 
To determine whether a conditional statement has been modified, we count the number of successor nodes in the control dependency graph (CDG) of the changed node. 
If the changed node has more than one successor, they are considered to be inside a conditional statement. 
We calculate the ratio of the affected nodes to the total number of nodes as the impact range of CDG.

The value of the affected range is calculated as the following formula, which merges the impact of CFG and DDG into a single value for further calculation.
\begin{equation}
    \label{eq:IR}
    IR=1+\sqrt{DDG\_impact}+\sqrt{CDG\_impact}
\end{equation}
where $DDG\_impact$ and $CDG\_impact$ are affected range from DDG and CFG respectfully. 
Since $DDG\_impact$ and $CDG\_impact$ are values between 0 and 1. 
And $IR$ is later calculated in the multiplicative form together with the other components. 
In order to avoid the case of $IR=0$, we make sure that this item is not less than 1. 
Moreover, to make the variation of $DDG\_impact$ and $CDG\_impact$ more significant, we calculated the square root of these two items.

\subsubsection{Normalization}

% After refinement, the values of each dimension have different value domains, which creates great trouble for the subsequent data fusion. 
% For example, in project alibaba/fastjson, cyclomatic complexity has a mean of $13.41$, median value of $3.0$, standard deviation of $78.81$, and maximum value of $6667$. 
% However, in the same project, the consideration of the integrity of the project of some functions is below 1. 
% These two types of data are entirely meaningless if calculated together without pre-processing. 
After the refinement process, the values of each dimension may have different value domains, which can make it difficult to combine the data for subsequent analysis. 
For example, in the project alibaba/fastjson, the cyclomatic complexity has a mean of $13.41$, median value of $3.0$, standard deviation of $78.81$, and maximum value of $6667$, which is from a function with thousands of lines of code.
On the contrary, in the same project, the consideration of the inter-function interaction of some functions is below 1, which is extremely small compared to the value of CC. 
These two types of data are not directly comparable and require pre-processing before they can be meaningfully combined.

% Therefore, we applied the Box-Cox transformation\cite{boxcox} on the refined data to convert the final data distribution into a curve that is approximate to normal distribution. 
% And then, we adjust this distribution to a distribution with a mean of 1 and a standard deviation of $\frac{1}{3}$. 
% Under this distribution, the probability of each item being negative is less than $0.0015$, and $0$ is used instead when negative values occur. 
% Data to be normalized includes cyclomatic complexity, Halstead Volume, percentage of comment, line of code of the function, function importance, DDG impact, and CDG impact. 
To address this issue, we apply the Box-Cox transformation~\cite{boxcox} on the refined data to convert the final data distribution into a curve that approximates a normal distribution. 
We then adjust this distribution to have a mean of 1 and a standard deviation of $\frac{1}{3}$. 
Under this distribution, the probability of each value being negative is less than $0.0015$, and a value of 0 is used instead of negative values. 
The data that is normalized includes cyclomatic complexity, Halstead Volume, percentage of comments, lines of code, the inter-function interaction of modified code segment, DDG impact, and CDG impact.

\subsection{Data Fusion}
% \zz{LACK CONNECTION}
% After refining each metric, we need to aggregate them into a single score. Firstly, we should integrate the complexity metrics into one single value. 
% Among the complexity metrics, complexity is positively correlated with the number of lines of code, Halstead volume, cyclomatic complexity, and negatively correlated with the percentage of comments. Therefore the complexity is defined as:
After refining each metric, we need to aggregate them into a single score. 
To do this, we first integrate the complexity metrics into a single value. 
Among the complexity metrics, there is a positive correlation between complexity and the number of lines of code, Halstead volume, cyclomatic complexity, and a negative correlation with the percentage of comments. 
Therefore, complexity is defined as:
\begin{equation}
    CM = \frac{1}{2}(LOC+CC+HV-PCom)+1
\end{equation}
where $LOC$ is line of code of changed function, $CC$ is cyclomatic complexity, $HV$ is Halstead volume and $PCom$ is percentage of comments. 
After integration, $CM$ has a distribution that has a mean of $2$, and a minimum of $1$.

% Next, we fuse the complexity with other metrics. The final score should satisfy the following characteristics.
% When a developer modifies, the contribution value is positively related to the amount of code being modified, the complexity, the consideration of the integrity of the project, and the impact of the change within the function, respectively. 
% Therefore, we use the following formula to define the contribution value in a function:
Next, we combine the complexity metric with other metrics. The final score should have the following characteristics:
when a developer makes modifications, the contribution value should have a positive relationship with the amount of code being modified, the complexity, and the inter-function and intra-function impact of the changed code segment.
Therefore, we use the following formula to define the contribution value for modifying a function in a commit:
\begin{equation}
    Score=\Delta_{AST}\cdot CM\cdot (IP+1)\cdot IR
\end{equation}
where $\Delta_{AST}$ is AST Difference, $CM$ is the complexity, $IP$ is the inter-function impact of the code segment from the call graph and $IR$ is the intra-function impact. We add $1$ to $IP$ and make sure that it is not smaller than 1. And the total contribution of a commit is the sum of all the contribution of modified functions:
\begin{equation}
    CValue = \sum_{i}^{set(f)} Score_{f_{i}}
\end{equation}

\section{Evaluation}

% In the evaluation part, we aim to answer the following research questions:
% \begin{enumerate}
%     \item How is the accuracy of \tool compared to the existing methods of measuring developer contribution on real-world projects?
%     \item How is the performance of \tool?
%     \item What are the potential applications of \tool?
% \end{enumerate}

The evaluation section of our paper aims to assess the effectiveness of \tool, in measuring the contribution of open-source developers. 
Specifically, we will address the following three research questions:
\begin{enumerate}
    \item How is the accuracy of \tool compared to the existing methods of measuring developer contribution on real-world projects?
    \item How is the performance of \tool?
    % \item \td{What are the behaviors of developers in open-source projects?}
    % 可以加解释说明
    % 
    \item What are the potential applications of \tool?
\end{enumerate}

\subsection{Experiment Setup}
\tool is not an effort estimation method but a contribution measurement method. We will not use the same accuracy metrics as effort estimation and compare to state-of-the-art methods of effort estimation.
\subsubsection{Baseline Selection} 

% \tool measures developer contributions at the semantic-level code changes, so the tools we compare use only code changes to measure developer contributions.

Our tool, \tool, assesses developer contributions at the semantic level by analyzing code changes. 
As a result, the tools we compare in this study also use code changes from the software version control system as input for measuring developer contributions.
We have chosen the following two methods for comparison in our study:

\noindent \textbf{Changed line of code (by git diff~\cite{gitdiff}).} 
This method is used in Github, which only count the number of changed line of code, including addition and deletion.
% Considering that the value of COCOMO is monotonically increasing with $KLOC$, which is the number of lines of changed code and we are using spearman correlation for evaluation, the result of COCOMO should be consistent with the changed lines of code.
% Since the value of COCOMO is directly proportional to $KLOC$, which represents the number of lines of changed code, and we are using spearman correlation for evaluation, the results obtained using COCOMO should align with those obtained using the number of changed lines of code.

\noindent \textbf{ELOC of Merico~\cite{merico}.} 
This method is proposed by Merico, a startup company by Yin et al.~\cite{Yin:EECS-2018-174,ren2018towards}, who proposed a method to quantify the contribution made by developers, to measure the efficiency of developers.
This method considers the number of changed AST nodes, the weight of each kind of node and edit type, and intra-function deduplication.

% COCOMO~\cite{boehm1995cost,boehm2008achievements} is a commonly used developing effort estimation model, which can be expressed by the following equation.
% \begin{equation}
%     E=a_{i}(KLoC)^{b_{i}}(EAF)
% \end{equation}
% where $a_{i}$ and $b_{i}$ are coefficients larger than zero, $KLoC$ is the estimated number of thousands of delivered lines of code for the project, $EAF$ is the effort adjustment factor that is larger than 0. 
% Since we used spearman correlation as an evaluation metric (in Section~\ref{sec:spearman}) which only considered the correlation of the rankings of several evaluation methods and did not consider the error, the performance of COCOMO is consistent with the method that only uses changed LoC.

We also considered COCOMO, a widely-used model for estimating development effort, as represented by the following formula, which value is only related to the lines of code. 
\begin{equation}
    \label{eq:cocoma}
    E=a_{i}(KLoC)^{b_{i}}(EAF)
\end{equation}
In this formula, $a_{i}$ and $b_{i}$ are coefficients greater than zero, $KLoC$ represents the estimated number of thousands of delivered lines of code for the project, and $EAF$ is the effort adjustment factor, which is also greater than zero. 
As we use Spearman correlation as an evaluation metric in Section~\ref{sec:spearman}, which only considers the correlation of the rankings of several evaluation methods and does not take into account the error, the result of COCOMO aligns with the method that counts only changed lines of code.

% Yin et al.~\cite{Yin:EECS-2018-174,ren2018towards} used DevRank and impact coding to measure the contribution of developers. 
% However, the weight of different types of impacting coding and the parameters in the Learning to Rank~\cite{liu2009learning} model was unavailable. 
% Therefore we were unable to reproduce the work in this study accurately.

% 先写选了哪几个，不选的部分精简一下，LoC和COCOMO一致，放在LoC后面
% \td{CONSIDER TO PUT IT BACKWARD}

There are other similar methods, but they are difficult to reproduce since some parameters are not publicly available.
Yin et al.~\cite{Yin:EECS-2018-174,ren2018towards} employed DevRank and impact coding to quantify developer contributions. 
However, the weights assigned to different types of impact coding and the parameters used in the Learning to Rank~\cite{liu2009learning} model were not disclosed, which prevented us from reproducing the results accurately in this study.
Bassi et al.~\cite{de2018measuring} and Chen et al.~\cite{chen2022code} used quality-based contribution metrics. 
However, the weights assigned to different types of quality metrics in their method are not publicly disclosed.
% For the quality-based contribution metric scheme, there is no open-source implementation of the study of the Bassi et al\cite{de2018measuring}. and the weights for the different types of quality metrics in their method are not open-source.
% The study by Chen et al\cite{chen2022code}. also does not have an open source implementation, and the state transition diagram proposed in the study only applies to Web projects and cannot be applied to other projects. 
% Therefore, both studies were not used for comparison.
% In the quality-based contribution metric scheme, there is no publicly available implementation of the study by Bassi et al.~\cite{de2018measuring}, and the weights assigned to different types of quality metrics in their method are not publicly disclosed. 
% Similarly, the study by Chen et al.~\cite{chen2022code} also lacks a public implementation and the state transition diagram proposed in the study is only applicable to web projects and cannot be extended to other types of projects. 
% As a result, neither study was included in our comparison.

% In addition, all the projects in the evaluation part are from open-source projects, but business value does not apply to open-source projects, so they are not adopted either. 
% Finally, we chose the following two methods to compare:
% Furthermore, all the projects in the evaluation section are from open-source projects, but the concept of business value is not applicable to open-source projects and therefore, it was not considered.

\subsubsection{Data Collection} 
% We conducted our evaluation on 1398 commits from 10 open-source projects with ground truth labeled by 10 skillful programmers. 
% The former 6 projects are from OSS-Fuzz~\cite{oss-fuzz}, which is a list of projects being fuzzed. 
% These projects have more maintaining operations in the evolution of history. 
% The latter 4 are smaller projects with simple functionality. 
% There is almost no operation of adding new features to the project in the modification history of these 4 projects.
We conducted our evaluation on 1398 commits from 10 open-source projects. 
The ground truth for these commits was labeled by 10 skillful programmers. 
The first six projects were selected from OSS-Fuzz~\cite{oss-fuzz}, which is a list of projects that are being fuzz-tested. These projects have both behaviors including adding features and maintaining existing code in their modification history. 
The last four projects are smaller projects with relatively simple functionality, and there are almost no new feature additions in their modification history.
% \zz{THE RULES TO SELECT THE 10 PROJECTS.}
% At least two programmers label each project to ensure the contribution score is not biased toward the preference of a single programmer.
% In cases where there was disagreement among the programmers labeling the data, we used the average of multiple people's labels to ensure fairness.
% Each programmer was asked to label the data according to his intuition by evaluating the contribution of the modified code to the project. 
% Modifications that did not change the functionality (e.g., formatting code only) received low scores. 
% Details of these projects are shown in Table~\ref{tab:project_accuracy}. 
% Data of the changed LoC is calculated from git diff~\cite{gitdiff}. 
% Data of the ELOC is from the Playground\cite{elocplayground} provided by Merico.
To ensure that the contribution scores are not biased towards the preferences of a single programmer, each project was labeled by at least two programmers. 
% In cases where there was disagreement among the programmers, we used the average of labels from multiple programmers to ensure fairness.
Each programmer was asked to label the data based on their intuition, by evaluating the contribution of the modified code to the project. 
Modifications that did not change the semantics (e.g., formatting code only) received low scores. The details of these projects are shown in 
Table~\ref{tab:project_accuracy}. The data for the number of changed lines of code was calculated using git diff~\cite{gitdiff} and the data for the estimated lines of code was obtained from the Playground~\cite{elocplayground} provided by Merico.
% \zz{ADD MORE ON HOW TO LABEL THE CONTRIBUTION, FOR EXAMPLE SCORE FROM 1-10 ETC.}

\begin{table}
\centering
\caption{Projects for accuracy evaluation}
\label{tab:project_accuracy}
\scriptsize
\begin{tabular}{lrrr}
\hline
\textbf{Project Name}         & \multicolumn{1}{l}{\textbf{Labelled}} & \multicolumn{1}{l}{\textbf{LoC}} & \multicolumn{1}{l}{\textbf{File}} \\ \hline
alibaba/fastjson              & 154                                           & 187k                             & 3,119                              \\
google/gson                   & 84                                            & 27k                              & 219                               \\
google/guice                  & 124                                           & 72k                              & 613                               \\
apache/httpcomponents-client  & 158                                           & 72k                              & 710                               \\
apache/httpcomponents-core    & 145                                           & 80k                              & 919                               \\
apache/rocketmq               & 135                                           & 169k                             & 1,552                              \\
apache/commons-cli            & 150                                           & 6,237                             & 52                                \\
apache/commons-release-plugin & 148                                           & 1,429                             & 18                                \\
apache/commons-exec           & 150                                           & 3,581                             & 52                                \\
apache/commons-ognl           & 150                                           & 20k                              & 305         \\                     
\hline
\end{tabular}
\end{table}

% \zz{DO NOT WRITE SENTENCE LIKE THAT.} 
% Since spearman correlation is used as an evaluation metric, each annotated data must be tagged by only one programmer. 
% It means that we cannot divide every single job among multiple programmers. 
% However, since labeling data is a very time-consuming task (about 100 commits per hour), the ground truth of each item we experimentally compare is labeled by only two programmers. 
% When there is a divergence in the data between the two developers, we will take the average as the result.
% We need to make sure that the results marked by different people are similar. 
% So, we let all programmers annotate 148 commits on apache/common-release-plugin and calculated the spearman correlation between the annotated results of different programmers. 
Since we are using Spearman correlation as an evaluation metric, each annotated data must be labeled by only one programmer. 
This means that we cannot divide the labeling task among multiple programmers. 
However, since the task of labeling data is time-consuming, taking about 80 commits per hour, the ground truth of each item we compared experimentally was labeled by only two programmers. 
In cases where there was disagreement among the programmers, a third developer will involve in the discussion and reach to an agreement.
To ensure that the results marked by different people were similar, we had all programmers annotate 148 commits on apache/common-release-plugin and calculated the Spearman correlation between the annotated results of different programmers to check for consistency.
\begin{figure}
    \centering
    \includegraphics[width=0.60\linewidth]{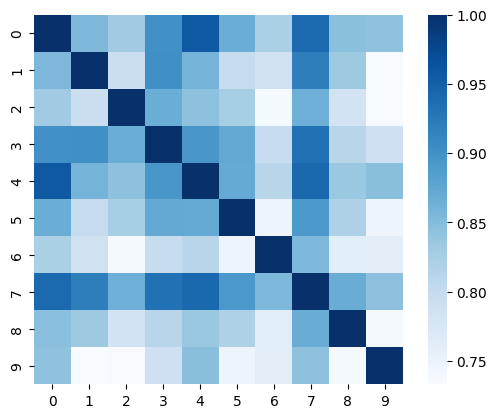}
    \caption{Correlation of manually labeled result}
    \label{fig:manually_labelled}
\end{figure}

% The result is shown in Figure~\ref{fig:manually_labelled}. According to the result, all 10 developers labeled data have a strong correlation with each other ($r_{s}>0.7$). 
% Thus, even though only 2 developers may label the data used in the experiment, it will not have a significant effect on the accuracy of the experiment.
The result is shown in Figure~\ref{fig:manually_labelled}. 
As can be seen, the labeled data from all 10 developers have a strong correlation with each other ($r_{s}>0.7$). 
This indicates that even though only 2 developers labeled the data used in the experiment, it will not have a significant impact on the accuracy of the experiment.

\subsubsection{Accuracy Metrics\label{sec:spearman}}

% To measure the similarity between the results obtained by the contribution measurement methods and the manually labeled results, we need a metric to measure the similarity or divergence. 
% Most of the previous studies used MAE (Mean Absolute Error), such as MSE (Mean Square Error). 
To assess the similarity between the results obtained by the contribution measurement methods and the manually labeled results, we need a metric to measure the similarity or divergence. 
Studies about effort estimation, which is similar to our topic, used metrics such as Mean Absolute Error (MAE) or Mean Square Error (MSE) to quantify the difference between the two sets of results.

% As an example, MSE calculates the mean value of the square of each error, as shown in the following equation:
As an example, MSE calculates the mean value of the square of each error, as shown in the following formula:
\begin{equation}
    MSE = \frac{1}{n}\sum_{i=1}^{n}(y_i - \hat{y_i})^2
\end{equation}
where $n$ is the number of samples, $y_i$ is the true value, and $\hat{y_i}$ is the predicted value.
From the formula, we can find three features of MSE. 
1) The minimum value of MSE is 0, but there is no upper bound. 
2) Assuming a fixed value of ground truth, the size and range of observations will significantly affect the value of MSE. 
3) MSE is not intuitive enough, and it is difficult to understand whether a value is a good enough result.

% Besides, it is difficult for developers to explain the differences in contribution scores\cite{Yin:EECS-2018-174,ren2018towards}, such as the difference between scores of 5 and 1. 
% The difference in scores can only represent the relative contribution size of two modifications. Therefore, we use spearman correlation as a measure. 
% Spearman correlation measures the monotonicity between two data distributions. It is calculated as shown in the following equation:
Additionally, it is challenging for developers to interpret the differences in contribution scores~\cite{Yin:EECS-2018-174,ren2018towards}, such as the difference between scores of 5 and 1. 
The difference in scores can only reflect the relative contribution size of the two modifications. 
Therefore, we use the Spearman correlation as a measure.
Spearman correlation measures the monotonicity between two data distributions. It is calculated using the following formula:
\begin{equation}
    r_{s} = \rho_{R(X),R(Y)} = \frac{cov(R(X),R(Y))}{\sigma_{R(X)}\sigma_{R(Y)}}
\end{equation}
where $R(X)$ is the rank variables of $X$, $cov(R(X),R(Y))$ is the covariance of the rank variables, $\sigma_{R(X)}$ is the standard deviation of the rank variables, $\rho_{R(X),R(Y)}$ is the Pearson correlation coefficient of the rank variables. 
This metric can address the three shortcomings of MSE mentioned above. 
1) spearman correlation has defined upper and lower bounds. 
2) The value of Spearman correlation is independent of the size and range of observations. 
3) The results of the Spearman correlation are intuitive. When $r_{s}=1$, it means that the ranking of the two data is completely positively correlated; when $r_{s}=0$, it means that the ranking of the two data is completely unrelated; when $r_{s}=-1$, it means that the ranking of the two data is completely negatively correlated.

\subsubsection{Experiement Environment} In all the experiments, we used the same experimental environment. 
We used a server with dual Intel(R) Xeon(R) 6248 CPUs and 188GB of RAM.

\subsection{Accuracy Evaluation (RQ1)}

% In RQ1, we aim to evaluate the accuracy of \tool in quantifying the contribution of each commit.
% In this chapter, two experiments are designed to evaluate the accuracy of \tool, and the reliability of manually annotated data.
In RQ1, our objective is to assess the accuracy of \tool in quantifying the contribution of each commit.
% \textbf{Experiment 1.} In Experiment 1, we try to evaluate the accuracy of our method and compare it with the baseline method. 
% Since ELOC Playground of Merico has errors in parsing some of the code, only commits without any file errors are considered when calculating the score corresponding to ELOC.
% From Table~\ref{tab:accuracy}, apart from google/gson, \tool performs better than the other two methods. 
% On these 10 projects, our accuracy exceeds LoC's method by $19.59\%$ and exceeds ELOC from Merico by $12.16\%$ on average. 
% 用两三个词描述一下Experiment
% \textbf{\yq{Accuracy Comparison} }
In this experiment, we evaluated the accuracy of our method and compare it with the baseline method. 
Since the ELOC Playground of Merico has errors in parsing some code, only commits without any file errors were considered when calculating the score corresponding to ELOC.
From Table~\ref{tab:accuracy}, it can be seen that, with the exception of google/gson, \tool performs better than the other two methods. 
On these 10 projects, our method outperforms the LoC-based methods, including COCOMO, by an average of $19.59\%$ and exceeds the ELOC method from Merico by an average of $12.16\%$.

\begin{table}[t]
\centering
\caption{Result of accuracy evaluation}
\label{tab:accuracy}
\scriptsize
\begin{tabular}{lrrr}
\hline
\textbf{Project}                       & \textbf{LoC} & \textbf{ELOC} & \textbf{\tool} \\ \hline
\textbf{alibaba/fastjson}              &  0.5319      &  0.5490       & 0.6210                           \\
\textbf{google/gson}                   &  0.5778      &  0.8223       & 0.7783                           \\
\textbf{google/guice}                  &  0.5841      &  0.5480       & 0.6935                           \\
\textbf{apache/httpcomponents-client}  &  0.4741      &  0.5161       & 0.5351                           \\
\textbf{apache/httpcomponents-core}    &  0.7525      &  0.7158       & 0.7935                           \\
\textbf{apache/rocketmq}               &  0.8649      &  0.7549       & 0.9009                           \\
\textbf{apache/commons-cli}            &  0.3685      &  0.5718       & 0.5736                           \\
\textbf{apache/commons-release-plugin} &  0.7062      &  0.5337       & 0.7084                           \\
\textbf{apache/commons-exec}           &  0.4947      &  0.6018       & 0.6458                           \\
\textbf{apache/commons-ognl}           &  0.5319      &  0.5490       & 0.6210                           \\ \hline
\end{tabular}
\end{table}

Compared to the LoC-based approach, the most significant improvement of $55.66\%$ was achieved on the project apache/commons-cli. 
This is because there are a large number of operations on this project that cannot be accurately identified by the LoC-based method, such as variable name modification. 
On http-components-core, rocketmq, and commons-release-plugin of Apache, the improvement is relatively small due to less worthless maintenance behavior and the already high LoC-based accuracy ($>70\%$).
Apart from the proportion of worthless change in the evolution history, other reasons would be discussed later.

% Since ELOC has errors in some of the modifications, we only use the part of the labeled data, which has correct results from ELOC, to calculate the correlation value. On project google/gson, our method was $5.35\%$ less than ELOC. 
% However, on this project, only 22 commits got the correct ELOC results in 84 commits without significant statistical value. 
% Furthermore, since the computational logic of ELOC is not fully open-sourced, we cannot analyze the reasons in detail for their accuracy.
As ELOC has errors in some modifications, we only used the part of the labeled data that had correct results from ELOC to calculate the correlation value. 
This means that the sample size of ELOC is smaller than the other methods.
On the project google/gson, our method was $5.35\%$ less accurate than ELOC. 
However, on this project, only 22 commits had correct ELOC results out of 84 commits, which does not provide a significant statistical value. 
Furthermore, since the computational logic of ELOC is not fully open-sourced, we cannot analyze the reasons for its accuracy in detail.

% In order to analyze which types of modifications would create a significant difference in results with the previous methods, we analyzed commits whose difference in ranking between the results of LoC, ELOC and \tool is greater than 40\%. 
% We have found the following change types.
In order to understand which types of modifications would result in significant differences in results when compared with the previous methods, we analyzed commits whose difference in ranking between the results of LoC, ELOC, and \tool is greater than $40\%$. We found the following change types:

% The first type contains some refactoring operations. 
% These operations involve changing variable names or method names, changing modifiers of properties or methods, moving methods, and extracting methods. 
% Operations of moving methods and extracting methods often move code from one place to another without producing much other modification. 
% They are recognized as move operations in the AST mapping algorithm. 
% Modifying modifiers only causes few AST nodes to be changed. 
% Operations that only modify variable names are blocked or given low weight when calculating AST changes. 
% All of these types of refactoring operations result in a large number of lines of code being modified. 
% However, there is no impact on the functionality of the program.
% 加点例子，如Github链接
% \td{CAN ADD IN MORE NUMBERS AND CODE EXAMPLE?}
The first type contains some simple refactoring operations\footnote{\url{https://github.com/alibaba/fastjson/commit/8d4ac6}}. 
These operations include changing variable names or method names, modifying the properties or methods modifiers, moving methods, and extracting methods. 
Commits of moving methods and extracting methods often move code from one place to another without producing much semantic modification. 
They are recognized as move operations in the AST mapping algorithm.
Modifying modifiers only causes a few AST nodes to be changed.
Commits that only modify variable names are blocked or given low weight when calculating AST changes. 
All of these types of refactoring operations result in a large number of lines of code being modified. 
However, such changes produce little semantic change and do not affect the operation of the program if not causing syntax errors.

% The second type is the modification of comments. 
% A comment helps the developer understand the program quickly, but it does not make any improvements to the functionality of the program. 
% When the change of AST change is calculated, it does not incorporate the comments into the AST. 
% Therefore, changing comments in the source code will not be counted as contributing to our approach.
The second type is the modification of comments\footnote{\url{https://github.com/alibaba/fastjson/commit/1c087e}}. 
Comments help developers understand the program quickly, but they do not make any semantic changes to the program. 
When calculating AST changes, comments are not counted into the AST. 
Therefore, changing comments in the source code will not be counted as contributing to our approach.

% The third category is for modifications of lengthy conditional statements. 
% Modifications in conditional statements affect the logic of the program, so we use the impact range to increase the contribution value of such modifications. 
% Besides, the number of AST nodes in a lengthy conditional statement is large and may even be higher than in a regular statement. 
% Such modifications would get a relatively high score in our method but would be no different from modifying just one line of code in an LoC-based method.
The third category is for modifications of lengthy conditional statements\footnote{\url{https://github.com/alibaba/fastjson/commit/3f28f5}}. 
Modifications in conditional statements affect the logic of the program, so we use the impact range to increase the contribution value of such modifications. 
Also, the number of AST nodes in a lengthy conditional statement is large and may even be higher than in a regular statement. 
Such modifications would receive a relatively high score in our method, but they would be no different from modifying just one line of code in an LoC-based method.

% The last type is caused by the inherent defects of the AST mapping algorithm. 
% Sometimes the AST mapping algorithm treats a tiny modification as a huge one. 
% Fan et al.\cite{fan2021differential} found that all existing AST mapping algorithms have limitations and do not accomplish accurate mapping in all situations. 
% The problems of AST mapping algorithms affect the robustness of the current method to some extent.
The last type is caused by the inherent defects of the AST mapping algorithm. 
Sometimes the AST mapping algorithm treats a minor modification as a significant one. 
Fan et al.\cite{fan2021differential} found that all existing AST mapping algorithms have limitations and do not accomplish accurate mapping in all situations. 
The problems of AST mapping algorithms affect the robustness of the current method to some extent.

\answerbox{\textbf{Answer to RQ1:} 
% The proposed method is better than the baseline in accuracy, especially on larger-scale projects. 
% In addition, we found that the manually labeled data does not always match the expectations of developers, and the labelers cannot take into account all the factors that developers are concerned about.
The proposed method is better than the baseline in accuracy, especially on large-scale projects. 
% Additionally, we found that the manually labeled data does not always match the expectations of developers and the labelers cannot take into account all the factors that developers are concerned about. 
We also conclude the types of code change that cannot be measured properly by prior works.
This highlights the need for more accurate and automated methods for measuring developer contributions.
}

\subsection{Performance Evaluation (RQ2)}

% Since our proposed method analyzes information from a different dimension, including global call graphs and other analyses with a high time cost, introducing more information will improve the accuracy of the analysis. 
% Undoubtedly the method based only on changed lines of code or AST difference would be much faster than ours. 
% So we still need to ensure that our analysis method's time consumption is acceptable. 
Our proposed method uses call graphs and other high-time-cost analyses, resulting in increased accuracy. 
While existing methods, such as those based solely on changes in lines of code or AST differences, have faster execution times, our method aims to strike a balance between analysis accuracy and time consumption to ensure its practical in real-world projects.
In practice, our method need not necessarily match the execution time of line-of-code-based approaches, but rather maintain an acceptable time consumption according to the frequency of code updates.

In this research question, we will evaluate whether the time cost of \tool is acceptable in real-world projects. 
1) How much performance improvement has our incremental dependency extraction tool had compared to running a dependency extraction tool on each version? 
2) How much time is needed on average when analyzing a commit? 
3) How often code is updated in popular projects? 
% \zz{TRY TO REORDER THE 3 EXPERIMENT. PUT 3 INTO FIRST.}
% 1和3并起来，放在2前；1和3较长，2稍微短一点；3缺少分析

\textbf{Time Cost. } 
% In Experiment 1, we calculated the time cost of \tool on projects of different sizes. 
% In this experiment, the analysis program was limited to 20 CPU cores. 
% It will take too much time to analyze these projects. 
% Therefore, in this experiment, projects with fewer commits were selected, and all the commits were analyzed. 
% The experimental results are shown in Table ~\ref{tab:performance_ourtool}.
In this experiment, we aimed to evaluate the time consumption of our tool, \tool, on real-world open-source projects of varying sizes.
Extracting the dependencies of a project is a time-consuming task, but it is also necessary for \tool. 
We have optimized Depends to allow incremental analysis to improve efficiency. 
In this experiment, we first compare the performance improvement of dependency extraction.
% In Experiment 3, we aimed to compare the time consumption of our incremental dependency extraction tool, IncrementalDepends, with Depends\cite{multilangdependsdepends}. 
IncrementalDepend was limited to using only 2 CPU cores, while Depends was run with its default configuration. 
We ran IncrementalDepends on several projects and compared the execution time with that of Depends on the last version of each project. 
In this experiment, we also analyze the overall time cost of \tool on the same projects.
Since the first five projects are too large and may take several days to complete the entire analysis, we have only selected the most recent 200 commits for analysis.

The results are shown in Table ~\ref{tab:performance_dependency}.
In the table, \textit{IDepends} is the time of IncrementalDepends, and \textit{D Time Per Commmit} is the time cost of IncrementalDepends per commit.
From the table, we can find that our tool IncrementalDepends only toke about 1.06 seconds on average which is $8\%$ time of Depends per commit. 
Being able to do incremental analysis, especially on large projects, is a great improvement of \tool.
Our analysis tool, \tool, had an average execution time of 83.39 seconds per commit across different projects. 
However, it should be noted that the project apache/commons-pool had a higher execution time compared to other projects due to the higher number of modified files per commit.
\begin{table*}
\centering
\caption{Performance analysis for dependency extraction}
\label{tab:performance_dependency}
\scriptsize
\begin{tabular}{lrrrrrrr}
\hline
\textbf{Project}                      & \multicolumn{1}{l}{\textbf{LoC (Java)}} & \textbf{Commit Number} & \multicolumn{1}{l}{\textbf{Depends}} & \multicolumn{1}{l}{\textbf{IDepends Time Per Commits$^{*}$}} & \multicolumn{1}{l}{\textbf{Total Time}} & \multicolumn{1}{l}{\textbf{Time Per Commit}} \\ \hline
\textbf{alibaba/fastjson}             & 186,876                          & 3,970 (200)                   & 48.35 seconds                        & 4.31 seconds & 0 day(s) 05:52:08 & 104.20 seconds                                           \\
\textbf{google/gson}                  & 27,498                           & 1,706 (200)                  & 6.11 seconds                                        & 1.04 seconds    & 0 day(s) 04:59:21                                     & 89.81 seconds                         \\
\textbf{google/guice}                 & 72,154                           & 2,017 (200)                  & 12.49 seconds                        & 1.85 seconds    & 0 day(s) 05:52:08                                     & 105.64 seconds                             \\
\textbf{apache/httpcomponents-client} & 72,304                           & 3,376 (200)                  & 12.85 seconds                        & 0.65 seconds    & 0 day(s) 05:15:44                                     & 93.82 seconds                             \\
\textbf{apache/httpcomponents-core}   & 80,065                           & 3,674 (200)                  & 13.59 seconds                        & 0.70 seconds    & 0 day(s) 05:30:33                                     & 99.17 seconds                             \\
\textbf{apache/commons-cli}   & 6,237                           & 1,176                   & 6.84 seconds                        & 0.17 seconds      & 1 day(s) 05:48:10                                     & 91.23 seconds                           \\
\textbf{apache/commons-release-plugin}   & 1,429                           & 653                   & 6.53 seconds                        & 0.12 seconds    & 0 day(s) 03:14:54                                     & 17.91 seconds                             \\
\textbf{apache/commons-exec}   & 3,581                           & 755                   & 1.66 seconds                        & 0.14 seconds      & 0 day(s) 08:00:51                                     & 38.21 seconds                           \\
\textbf{apache/commons-ognl}   & 20,422                           & 809                   & 18.73 seconds                        & 0.50 seconds      & 0 day(s) 16:22:03                                     & 72.83 seconds                           \\
\textbf{apache/commons-pool}   & 15,405                           & 2,526                   & 5.17 seconds                        & 1.08 seconds      & 3 day(s) 12:55:53                                     & 121.04 seconds                           \\ \hline\\
\end{tabular}
\leftline{$^{*}$ Time for IncrementalDepends per Commit.}
\end{table*}

\textbf{A study of maintenance frequency of open-source software.} 
% In Experiment 2, we analyzed how many commits were made on average per day for popular open-source projects. 
% Since open-source projects do not always update every day, in this experiment, only days when changes were committed were taken into account. 
% By calculating the average number of commits per day, combined with the results in Experiment 1, we can conclude whether \tool can be used in practice. 
% In this experiment, we filtered $3108$ popular projects written in Java from GitHub and Maven Central\cite{maven}, categorized according to their sizes (number of lines of Java code). 
% The distribution of their daily commit numbers was counted, and the specific results are shown in Table~\ref{tab:commit_day_distrib}.
In this experiment, we aimed to evaluate the feasibility of our tool, \tool, in practice by analyzing the frequency of commits on popular open-source projects. 
Only the days on which changes were committed were taken into account, and the average number of commits per day was calculated. 
By combining this data with the results from Experiment~1, we can determine the practicality of using the \tool.

We randomly selected 3,108 popular Java projects from GitHub and Maven Central \cite{maven}, grouped by size (number of lines of Java code). The distribution of daily commit numbers was analyzed, and the results are presented in Table~\ref{tab:commit_day_distrib}.
In this table, the first column of the table represents different project sizes, measured by the number of lines of code.
From the results, it can be observed that even among projects of varying sizes, in more than $98\%$ of cases, the number of commits per day does not exceed 20. 
This is a relatively low number of commits that need to be analyzed, and even when considering the project with the highest execution time, the performance of \tool is practical for a maintenance frequency of up to 20 commits per day.
\begin{table}
    \centering
    \caption{Commit number per day on different projects}
    \scriptsize
    \begin{tabular}{lrrrr}
    \hline
    \textbf{LoC (Java)}    & \textbf{1-5} & \textbf{5-10} &\textbf{10-20} &\textbf{20+} \\ \hline
    \textbf{0-1k}   & 88.96\%                                     & 8.75\%                                      & 1.97\%                                       & 0.31\%                                     \\ 
    \textbf{1-5k}   & 85.11\%                                     & 11.46\%                                      & 2.88\%                                       & 0.54\%                                     \\ 
    \textbf{5-10k}  & 83.72\%                                     & 12.38\%                                      & 3.33\%                                       & 0.57\%                                     \\ 
    \textbf{10-50k} & 80.91\%                                     & 13.96\%                                      & 4.24\%                                       & 0.89\%                                     \\ 
    \textbf{50k+}   & 76.57\%                                     & 16.44\%                                      & 5.54\%                                       & 1.45\%                                     \\ \hline
    \end{tabular}
    \label{tab:commit_day_distrib}
\end{table}

\answerbox{\textbf{Answer to RQ2: }
% Although our approach has a larger time cost compared to line-of-code-based analysis, after calculating the maintenance frequency of open-source software and our average time cost, it shows that our tool's performance is acceptable.
While our approach may have a longer execution time compared to line-of-code-based analysis, upon considering the maintenance frequency of open-source software and the average time cost \tool, it can be concluded that the time cost of our tool is practical in real-world projects.
}

\subsection{Application (RQ3)}

% In RQ3, we try to discover the application scenarios of our approach. 
% Here, we try to find some unusual developers with our method. 
% Some developers, in order to get a higher place in various leaderboards (such as GitHub contributors of a specific project) or better performance awards in the company, often submit some worthless commits to the software repository. 
% These commits may be just fine-tuning non-core components of the project, maybe reformatting code, modifying documentation, etc. 
% Such changes do not require the developer to have in-depth knowledge of the project. Open-source projects do not necessarily benefit from such changes either. 
In RQ3, we investigate potential applications in real-world open-source projects for \tool. 
% In this RQ, we focused on identifying developers who may submit worthless commits in an effort to improve their standing on leaderboards (e.g. GitHub contributors of a specific project) or to earn awards within a company. 
Leaderboards, which are commonly used to show the developers with more frequent activities, and even used to rank the contribution of developers.
In this RQ, we want to verify whether the position of developers on the contribution leaderboard of GitHub is reasonable.
We analyzed the percentage of commits by each developer, and their percentage of \tool. 
For those whose \tool and number of commit have a significant gap, i.e. a reasonable threshold of the ratio of the percentage of commit number and \tool is used to find the gap, we labeled the change types in these commit and compare the proportion of each type of change between these two groups of developers.
% Such commits may include cosmetic changes, like adjustment of code style, or modifications of documentation, which do not require a deep understanding of the project and provide little benefit to the open-source project. 
% Then, we compared the code changes submitted by such developers with those submitted by other developers.
% Our method can assist in identifying such behavior.

% We analyzed 174 open-source Java projects from GitHub, identified developers based on the emails of commit operations in the git repository, and calculated the contribution share of each developer. 
% In order to discover the behavior of the same developer in different projects, the same developers from different projects are not merged. 
% In addition, we also calculated the share of the number of commits per developer, which will be used in the contributor ranking of this project on GitHub. 
% We compared the two ratios and filtered out developers whose commit count ratio was greater than $1\%$ and whose contribution value accounted for less than $20\%$ of the commit ratio, calling them developers with an inflated number of commits.
We conducted an analysis of 174 open-source Java projects from GitHub. 
We identified developers based on the emails associated with commit operations in the Git repository and calculated the contribution proportion for each developer in a project. 
% In order to understand the behavior of individual developers across different projects, we did not merge data for developers found in multiple projects. 
Furthermore, we calculated the proportion of commits per developer, which we used to determine the contributor ranking of the project on GitHub. 
By comparing these two proportions, we were able to filter out developers who had an inflated number of commits, i.e. commit number proportion greater than $1\%$ and contribution value proportion less than $20\%$ of the commit number proportion.

% We found 2282 developers with an inflated number of commits on 174 projects. 
% Among these, 2050 developers did not conduct syntax contributions to the Java code. 
% These changes from developers included, but were not limited to, build scripts, version numbers, comments and documentation, code format, etc. 
% The remaining 232 developers contributed far less than their share of commits. 
% In addition to this, 103 bots were found based on username. Most of these bots are Github Dependabot, mainly used to update dependent package versions, etc.
We identified 2,282 developers with inflated commit numbers across 174 projects. 
Of these, 2,050 did not make any syntax contributions to the Java code. 
These changes were primarily related to editing build scripts, version numbers, comments and documentation, code style, etc. 
The remaining 232 developers made significantly less contribution value than their proportion of commits suggested. 
Additionally, we identified 103 bots, most of which were Github Dependabot, which were mainly used to update dependent package versions.

\begin{figure}
    \centering
    \includegraphics[width=\linewidth]{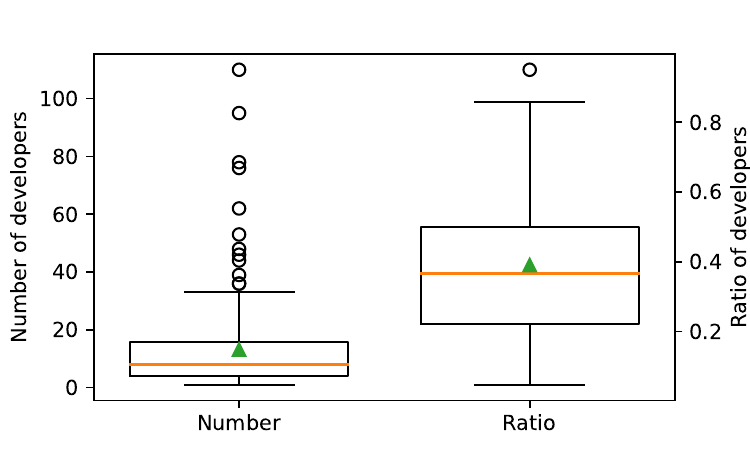}
    \caption{Developers with inflated commit number}
    \label{fig:boxplot}
\end{figure}

% As shown in Figure~\ref{fig:boxplot}, we show the results of our analysis on 174 open-source projects from GitHub. The left boxplot figure represents the number of developers who have an unusually high number of commits in different projects, and the right boxplot figure represents the proportion of these developers among all developers. From the figure, we can see that in these open-source projects, the average number of abnormal contributors per project is 13, with a median of 8; the average proportion is 39.01\%, and the median is 36.14\%.
Our analysis of 174 open-source projects from GitHub is shown in Figure~\ref{fig:boxplot}. 
The left boxplot illustrates the number of developers with an abnormal number of commits across different projects, while the right boxplot shows the proportion of these developers among all contributors. 
The figure demonstrates that in these open-source projects, the average number of developers who did not provide sufficient contribution per project is 13, with a median of 8. 
The average proportion of these contributors is $39.01\%$, with a median of $36.14\%$.

For the two categories of developers with a normal and inflated number of commit levels, we randomly sampled 200 commits submitted from each category and manually analyzed the modification content. 
As shown in Table \ref{tab:change_type}, developers with an inflated number of commits rarely proposed modifications that affected the semantics of the code. 
They focused more on improving documentation and resource files, modifying CI/CD scripts, and updating software dependencies. 
In contrast, developers with a higher ratio of contributions to commit numbers focused more on modifications to the code itself, including implementing new features, fixing errors, and updating versions. 
They paid less attention to resource files, dependency versions, etc. 
The latter were more likely to be core contributors to open-source projects.
\def\mybarC#1{#1
   {\color{black}\rule{\fpeval{#1/\percentscale*\barwidth} cm}{\barheight}\color{gray!30}\rule{\fpeval{(\percentscale-#1)/\percentscale*\barwidth} cm}{\barheight}} 
}
\newcommand{\barwidth}{0.7} % cm max bar widths
\newcommand{\barheight}{4pt} % height of each bar
\newcommand{\percentscale}{200} % max scale for percent bars
\newcommand{\degscale}{100} % max scale for degree bars

\begin{table}[t]
\centering
\caption{Change type of sampled 400 commits}
\label{tab:change_type}
\scriptsize
\begin{tabular}{lrr}
\hline
\textbf{Change Type}                        & \textbf{Inflated}       & \textbf{Normal} \\ \hline
\textbf{Add Empty Files} & \mybarC{2} & \mybarC{0} \\
\textbf{Semantic Changes} & \mybarC{7} & \mybarC{50} \\
\textbf{Rename Variables or Methods} & \mybarC{3} & \mybarC{5} \\
\textbf{Minor Semantic Changes} & \mybarC{20} & \mybarC{49} \\
\textbf{Change Build Scripts} & \mybarC{29} & \mybarC{22} \\
\textbf{Update Documentation} & \mybarC{27} & \mybarC{19} \\
\textbf{Change Dependency Version} & \mybarC{58} & \mybarC{10} \\
\textbf{Change Resource Files} & \mybarC{19} & \mybarC{5} \\
\textbf{Delete Files} & \mybarC{3} & \mybarC{4} \\
\textbf{Empty Commit} & \mybarC{5} & \mybarC{0} \\
\textbf{Non-Java Code} & \mybarC{2} & \mybarC{0} \\
\textbf{Format Code} & \mybarC{8} & \mybarC{5} \\
\textbf{Change Comments} & \mybarC{6} & \mybarC{11} \\
\textbf{Remove Redundant Code} & \mybarC{1} & \mybarC{1} \\
\textbf{Update CI/CD Scripts} & \mybarC{18} & \mybarC{5} \\
\textbf{Update Version} & \mybarC{16} & \mybarC{31} \\
\textbf{Move Files} & \mybarC{0} & \mybarC{2} \\

\hline
\end{tabular}
\end{table}

\begin{figure}
\centering
\begin{minipage}[t]{0.40\linewidth}
\centering
\includegraphics[width=\linewidth]{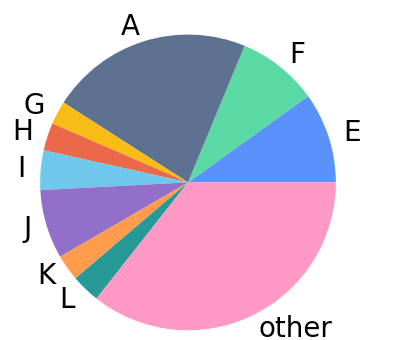}
\end{minipage}
\begin{minipage}[t]{0.40\linewidth}
\centering
\includegraphics[width=\linewidth]{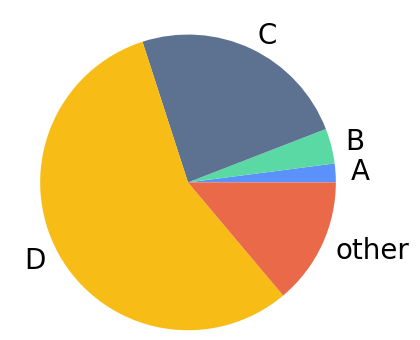}
\end{minipage}
\caption{Percentage of contribution by commit and contribution}
\label{fig:developer_share}
\end{figure}

% The following case is from the open-source project LibrePDF/OpenPDF on GitHub. 
% As shown in Figure~\ref{fig:developer_share}, in the first figure, according to the number of commits, the core developer should be A. 
% In the first graph, it occupies $22.17\%$ of the pie chart, but in the second graph, it occupies only $2.04\%$. 
% If we look at its commits on GitHub, many of its commits are updating version numbers, merging reviewed pull requests, and updating dependencies. 
% Even though it has the highest number of commits in the project, it does not contribute as much to the project as other developers. 
% E in the diagram is a bot that, by name, is a GitHub Dependabot, which also has submitted a large number of commits, but all of these commits are modifying dependencies of the project. 
% As shown in the image on the right, it makes almost no contributions. 
% Also, B, C, and D are all project developers, and D submitted the most lines of code. 
% In the left figure, they do not appear because the number of commits is too small. 
% B submits fewer lines of code and fewer commits than A, but in the right figure, we find that B has a more significant contribution than A.
The following case is from the open-source project LibrePDF of OpenPDF on GitHub. 
As shown in Figure~\ref{fig:developer_share}, developer A appears to be the core developer according to the number of commits, occupying $22.17\%$ of the pie chart in the first figure. 
However, in the second figure, its contribution share is only $2.04\%$. 
When looking at the commits by A on GitHub, many of them are related to updating version numbers, merging reviewed pull requests, and updating dependencies, even though it has the highest number of commits in the project, it does not contribute as much as other developers.
Developer E in the figure is a bot, identified as GitHub Dependabot, which has also submitted a large number of commits, but these commits are all related to modifying the dependencies of the project. 
Therefore, as shown in the second figure, it makes almost no contributions. 
Developers B, C, and D are all project developers, and developer D produced the most contribution. 
In the first figure, they do not appear because their number of commits is too small. 
However, Developer B submits fewer lines of code and fewer commits than A, but as seen in the second figure, B has a more significant contribution than A.

%\zz{PLEASE DO NOT USE REAL NAMES?}

\answerbox{\textbf{Answer to RQ3:} 
% After analyzing a specific size of open source projects, we found that there are specific application scenarios for our approach to finding developers whose commit numbers do not match their contributions. 
After analyzing 174 open-source projects, we discovered that our approach to identifying developers whose commit numbers do not align with their contributions is useful in real-world open-source projects.
}
\section{Threats to Validity}

In RQ1, where we compare the accuracy of several schemes to manually labeled data, the manually labeled data may be not accurate enough, which has two reasons. 
First, the individuals who annotated the data were not involved in the development of the projects they were annotating, which eliminates potential bias but may also lead to a lack of understanding of the content and the difficulty of the modifications. 
Second, developers do not always have the same focus when writing code and reviewing code. 
The views of reviewers of the revisions do not always accurately reflect what the developers were thinking.
These two factors could potentially compromise the accuracy of RQ1.

\section{Conclusion}

% In conclusion, we proposed \tool, a method for measuring the contribution of open-source software developers. In the proposed method, we combined data from four aspects of the software repository aspects. Our method is evaluated by comparing it to the previous method on 10 projects containing 1398 commits. The data indicate that \tool outperforms the previous method by an average of $19.59\%$. Moreover, we notice that manually labeled data may not accurately reflect the concern of developers when developing. We investigate the robustness and performance of the proposed method to assure its applicability in the real world. Finally, we analyzed the contributions of developers on 276 open-source projects and discovered 3155 developers who did not make enough contributions.
In conclusion, our research proposed \tool, a method for accurately measuring the contributions of open-source software developers. 
By combining data from four aspects of the software repository, our method was able to outperform previous methods by an average of $19.59\%$ when evaluated on 10 projects containing 1,398 commits. 
% Additionally, we acknowledged that manually labeled data may not always reflect the true concerns of developers and thus investigated the robustness and performance of \tool to ensure its applicability in real-world scenarios. 
By evaluating the time cost of \tool on 10 popular open-source projects, \tool takes 83.39 seconds per commit on average and it is practical in the real world.
Through our analysis of 174 open-source projects, we also discovered 2,282 developers who were not making enough contributions.
On all 174 projects, developers who did not contribute enough had an average of 13 people per project and a proportion of $39.01\%$.
Overall, our research provides a valuable tool for measuring the contributions of open-source developers and identifying those who may need additional attention.
% \zz{WILL WE OPEN SOURCE OUR TOOL AND OUR TESTING DATA???}

\section*{Acknowledgement}

This research/project is supported by the National Research Foundation Singapore and DSO National Laboratories under the AI Singapore Programme (AISG Award No: AISG2-RP-2020-019), the National Research Foundation, Singapore, and the Cyber Security Agency under its National Cybersecurity R\&D Programme (NCRP25-P04-TAICeN). 
Any opinions, findings and conclusions or recommendations expressed in this material are those of the author(s) and do not reflect the views of National Research Foundation, Singapore and Cyber Security Agency of Singapore.

% \input{texs/data.tex}
\bibliographystyle{IEEEtran}
\bibliography{ref}

\end{document}